\documentclass{article}
\usepackage{spconf,amsmath,epsfig}
\usepackage{amsfonts}
\usepackage{amssymb}
\usepackage{cite}
\usepackage{color}
\ninept

\newtheorem{theorem}{Theorem}

\newtheorem{formulation}{Formulation}

\newtheorem{problem}{Problem}

\title{Transceiver Pair Designs for Multiple Access Channels under Fixed Sum Mutual Information using MMSE
Decision Feedback Detection}
\name{Wenwen Jiang, Jian-Kang Zhang and Kon Max Wong}
\address{Department of Electrical and Computer Engineering, \\McMaster
University, Hamilton, Ontario, Canada. \\ Emails:
jiangw8@mcmaster.ca; (jkzhang, wong)@mail.ece.mcmaster.ca}
\begin{document}
%
\maketitle
\begin{abstract}
In this paper, we consider the joint design of the transceivers
for a multiple access Multiple Input and Multiple Output (MIMO)
system having Inter-Symbol Interference (ISI) channels. The system
we consider is equipped with the Minimum Mean Square Error (MMSE)
Decision-Feedback (DF) detector. Traditionally, transmitter
designs for this system have been based on constraints of either
the transmission power or the signal-to-interference-and-noise
ratio (SINR) for each user. Here, we explore a novel perspective
and examine a transceiver design which is based on a fixed sum
mutual information constraint and minimizes the arithmetic average
of mean square error of MMSE-decision feedback detection. For this
optimization problem, a closed-form solution is obtained and is
achieved if and only if the averaged sum mutual information is
uniformly distributed over each 
active subchannel. Meanwhile, the mutual information of the
currently detected user is uniformly distributed over each
individual symbol within the block signal of the user, assuming
all the previous user signals have been perfectly detected.
\end{abstract}
\section{Introduction}
\label{sec:intro}

We consider a block-based synchronous multiple access frequency
selective MIMO channel in which the users' data sequences are
precoded separately and are transmitted over distinct ISI
channels. If we denote the signal vector for the $k$th user as
${\mathbf x}_k$, $k=1,\cdots,K$, then, the received signal vector
can be represented by
\begin{eqnarray}\label{precoded multiple access channel}
    {\mathbf y}=\sum_{k=1}^K{\mathbf H}_k{\mathbf T}_k{\mathbf
    x}_k+\boldsymbol\xi
\end{eqnarray}
where ${\mathbf H}_k$ is a $P\times M$ block Toeplitz tall channel
matrix corresponding to zero-padded modulation or an $M\times M$
square block diagonal channel matrix corresponding to Discrete
Multiple Tone (DMT) modulation~\cite{chow91, ruiz,
ginis02,iterative_waterfilling} for the $k$th user, and ${\mathbf
T}_k$ is an $M\times N_k$ precoder matrix for the $k$th user, and
$\boldsymbol \xi$ is the zero-mean Gaussian noise vector. Our task
in this paper is to obtain an optimum design for all the $K$
transceivers for such a system using MMSE-decision feedback
detection.


In recent years, for the single-user MIMO system using linear
receivers, there exist solutions to a large number of precoder
design problems~\cite{yang94, palomar03}, including the
maximization of information rate~\cite{scaglione} and of
SNR~\cite{giannakis99}, the minimization of the mean squared
error~\cite{giannakis99} and of bit error probability for both
zero-forcing~\cite{ding} and MMSE
equalization~\cite{davidson04_block}. For a multiuser system using
linear MMSE receivers, the joint design of transceivers minimizing
the total MSE was efficiently implemented by solving a convex
optimization problem~\cite{Jorswieck03,multiuser_linear}. MIMO
systems having DF detectors have also been studied. For
single-user systems equipped with ZF-DF~\cite{jkzhang_qrs1} and
MMSE-DF~\cite{singleuser_dfe,jkz-isit05} detectors, closed-form
optimal transceivers have been obtained using the equal diagonal
QRS decomposition of a matrix~\cite{jkzhang_qrs1}. However, these
solutions for the single-user system cannot be directly
generalized to a multiuser case. 

In this paper, we examine the optimum design of transceivers of a
MIMO system using DF detectors in a multiuser environment. We note
that all existing transceiver designs thus far have been pursued
based on solving some optimization problems subject to some power
constraints or to some SINR constraints on each user. In this
paper, we explore a novel perspective and examine the transceiver
design by minimizing the mean square error for $K$ users subject
to a fixed sum mutual information constraint. Here, we focus our
consideration on a block-by-block ISI multiple access MIMO system
employing the MMSE-DF detector.

{\bf{Notation}}: Matrices are denoted by uppercase boldface
characters (e.g., ${\mathbf A}$), while column vectors are denoted
by lowercase boldface characters (e.g., ${\mathbf b}$). The
$(i,j)$-th entry of ${\mathbf A}$ is denoted by $A_{i,j}$. The
$i$-th entry of ${\mathbf b}$ is denoted by $b_i$. The columns of an
$M\times N$ matrix ${\mathbf A}$ are denoted by ${\mathbf a}_1,
{\mathbf a}_2, \cdots, {\mathbf a}_N$. Notation ${\mathbf A}_k$
denotes a matrix consisting of the first $k$ columns of ${\mathbf
A}$, i.e., ${\mathbf  A}_k=[{\mathbf a}_1, {\mathbf a}_2, \cdots,
{\mathbf a}_k]$. By convention, ${\mathbf A}_0=1$. The $j$-th
diagonal entry of a matrix ${\mathbf A}$ is denoted by $[{\mathbf
A}]_j=A_{j}$. The Hermitian transpose of ${\mathbf A}$ (i.e., the
conjugate and transpose of ${\mathbf A}$) is denoted by ${\mathbf
A}^H$.

\section{The Inverse Water-filling Solution}
\label{sec:inverse}

We first provide the inverse water-filling solution on which our
transceiver design is based. Consider a general single-user MIMO
system such that
\begin{eqnarray}\label{single_model}
    {\mathbf y}={\mathbf H}{\mathbf T}{\mathbf x}+\boldsymbol\xi
\end{eqnarray}
where ${\mathbf H}$ is an $P\times M$ complex matrix, ${\mathbf
T}$ is an $M\times N$ matrix and ${\boldsymbol\xi}$ is an $P\times
1$ Gaussian noise vector with a covariance matrix ${\mathbf\Xi}$.
It is well known that if the channel matrix ${\mathbf H}$
in~({\ref{single_model}) is known at both the transmitter and the
receiver, then, the Gaussian mutual information of
model~(\ref{single_model}) is given by ${\mathcal
I}=\log\det({\mathbf I}+{\mathbf T}^H{\mathbf
H}^H{\mathbf\Xi}^{-1}{\mathbf H}{\mathbf T})$~\cite{cover}. Thus,
subject to a power constraint ${\rm tr}({\mathbf T}^H{\mathbf
T})\le p$, the channel capacity is achieved when the optimal
transmitter $\widetilde{\mathbf T}$ is the water-filling
solution~\cite{brandenburg74}. In the inverse water-filling
problem, instead of constraining the power, the total transmission
power is minimized subject to a fix Gaussian mutual information,
which can be stated as
\begin{problem}\label{prob:throughput} Find an optimal transmitter
    ${\mathbf T}$ such that
    \begin{eqnarray}
        \min_{\mathbf T}&\text{tr}({\mathbf T}^H{\mathbf T})\nonumber\\
        {\rm s.t.}&\log\det({\mathbf I}+{\mathbf T}^H{\mathbf H}^H{\mathbf
        H}{\mathbf T})={\mathcal I}\nonumber
    \end{eqnarray}
\end{problem}
where, for simplicity, we have assumed the channel noise is
uncorrelated. If the eigenvalue decomposition of ${\mathbf
H}^H{\mathbf H}$ is ${\mathbf U}{\mathbf\Lambda}{\mathbf U}^H$
with eigenvalues $\lambda_i$ arranged in non-increasing order,
then, the optimal solution to the Problem~\ref{prob:throughput} is
given by ${\mathbf T}={\mathbf U}_r{\boldsymbol
\Gamma}^{1/2}{\mathbf S}$, where ${\mathbf S}$ is an arbitrary
unitary matrix, ${\bf U}_r$ consists of the first $r$ columns of
the unitary matrix ${\bf U}$, ${\mathbf\Gamma}={\rm
diag}(\gamma_1, \gamma_2, \cdots, \gamma_{r})$ with each
$\gamma_n$ determined by\vspace{-0.2cm}
\begin{eqnarray}\label{loading}
    \gamma_n=\left(\frac{2^{{\mathcal
    I}}}{\prod_{i=1}^{r}\lambda_i}\right)^{1/r}-\lambda_n^{-1}
\end{eqnarray}
and $r$ is the largest integer index, not exceeding $M$, for the
eigenvalues such that \vspace{-0.2cm}
\begin{eqnarray}
    \lambda_{k}>\left(\frac{\prod_{i=1}^{r}\lambda_i}{2^{{\mathcal
    I}}}\right)^{1/r}\qquad {\rm for}\ k=1,2,\cdots,r
\end{eqnarray}
Similar research work to~Problem~\ref{prob:throughput} can be found
in~\cite{jorswieck07,cioffi06}

\section{JOINT DESIGN OF TRANSCEIVER FOR MULTIPLE ACCESS CHANNELS}
\label{sec:transiver_design}

With the inverse water-filling solution in mind, we now consider
the joint design of the transceivers for a multiple access ISI
MIMO system equipped with the MMSE-DFE. The goal of our design is
to minimize the arithmetic MSE for the $K$ users subject to a
fixed sum mutual information constraint.

\subsection{Problem Statement and Formulation}\label{sec:mmse}

In general, for a DF receiver, signals are detected in the reverse
order of the user index, i.e., we first detect the signal from
User $K$, then that from User $K-1$, and so on. Based on this
detection order, we thus re-write the received signal
of~(\ref{precoded multiple access channel}) as
\begin{eqnarray}\label{MMSE-DFdetection}
    {\mathbf y}-\hspace{-0.2cm}\sum_{i=k+1}^{K}\hspace{-0.2cm}{\mathbf
    H}_{i}{\mathbf T}_{i}{\mathbf
    x}_{i}\hspace{-0.2cm}&=&\hspace{-0.2cm}{\mathbf H}_k{\mathbf
    T}_k{\mathbf
    x}_{k}+\underbrace{\hspace{-0.2cm}\sum_{\ell=1}^{k-1}{\mathbf
    H}_{\ell}{\mathbf
    T}_{\ell}{\mathbf x}_\ell+\boldsymbol\xi}_{\boldsymbol\zeta_{k}}\nonumber\\
    \hspace{-0.2cm}&=&\hspace{-0.2cm}{\mathbf H}_k{\mathbf T}_k{\mathbf x}_{k}
    +{\boldsymbol\zeta_{k}},~~~ k=K, \cdots,1.
\end{eqnarray}
where ${\boldsymbol\zeta_k}$ is the $k$th interference-plus-noise
vector. In Eq. (\ref{MMSE-DFdetection}), the MMSE-DFE is used to
detect ${\mathbf x}_k$ from the received signal ${\mathbf y}$ by
successively cancelling the previously detected user signals.

Let ${\mathbf B}$ and ${\mathbf F}$ be the feedback and feedforward
matrices of the MMSE-DFE respectively. Exploiting the orthogonality
principle~\cite{proakis} and using the Matrix Inversion
Lemma~\cite{matrix_formula} leads the optimum ${\mathbf F}_{MMSE,k}$
which results in the error covariance matrix for User $k$ being
given
by~\cite{singleuser_dfe,cioffi_mmse_dfe,cioffi_dfe,varanasi_dfe}
\begin{eqnarray}\label{error_covariance}
{\mathbf A}_k &=&E[{\mathbf e}_{k}{\mathbf
e}_{k}^H]\nonumber\\
&=&{\mathbf W}_k({\mathbf
J}_{k})^{-1}({\mathbf W}_k)^H\nonumber\\
&=&{\rm diag}([{\mathbf R}_k]_1^{-2},[{\mathbf
R}_k]_2^{-2},\cdots,[{\mathbf R}_k]_{N_k}^{-2}).
\end{eqnarray}
where
\begin{eqnarray}
{\mathbf J}_{k}&=&{\mathbf I}+({\mathbf H}_k{\mathbf
T}_k)^H({\boldsymbol\Sigma}_k)^{-1}{\mathbf H}_k{\mathbf T}_k\nonumber\\
{\boldsymbol\Sigma}_k&=&E[{\boldsymbol\zeta}_{k}{\boldsymbol\zeta}_{k}^H]={\mathbf
I}+\sum_{\ell=1}^{k-1}{\mathbf H}_{\ell}{\mathbf T}_{\ell}({\mathbf
H}_{\ell}{\mathbf T}_{\ell})^H \nonumber\\
&&{\rm for}\ k=1,2,\cdots,K\ {\rm and}\ {\boldsymbol\Sigma}_{1}={\mathbf I}\nonumber\\
{\mathbf W}_k&=&{\mathbf B}_k+{\mathbf I}\nonumber\\
\end{eqnarray}
and $\mathbf R$ is the upper triangular matrix obtained by the
$QR$-decompostion of $\mathbf J^{1/2}$. We note that ${\mathbf W}_k$
is an upper-triangular matrix with unit diagonal entries. If we
define the average MSE of the $K$ users of the successive
cancellation detector as
\begin{eqnarray}\label{mmse-dfe-error}
{\mathcal E}
\hspace{-0.2cm}&\triangleq
&\hspace{-0.2cm}\frac{1}{N}\sum_{k=1}^K{\rm tr}\left({\rm
E}[{\mathbf e}_k{\mathbf e}_k^H]\right)=\frac{1}{N}\sum_{k=1}^K{\rm
tr}\left({\mathbf A}_k\right)
\end{eqnarray}
where $N=\sum_{k=1}^{K}N_k$, our optimization problem can be
formally stated as follows:
\begin{problem}\label{prob:mmse-error}
Let ${\rm rank}({\mathbf H}_k)=L_k$, $k=1, 2, \cdots, K$. Then,
given $K$ non-negative integers $N_1, N_2, \cdots, N_K$ with
$N_k\le L_k$, find the matrix sequence $\{{\mathbf T}_k\}_{k=1}^K$
such that
\begin{enumerate}
\item the MMSE for the $K$ users of the MMSE-DF detection is first
minimized, subject to a fixed sum mutual information constraint,
i.e.,
\begin{eqnarray}
\{\widetilde{\mathbf T}_k\}_{k=1}^K=\arg\min~ {\mathcal
E}
\end{eqnarray}
s.t.
\begin{eqnarray}\label{sum mutual information constraint}
{\mathcal I}=\log\det\left({\mathbf I}+\sum_{k=1}^K{\mathbf
H}_k{\mathbf T}_k{\mathbf T}_k^H{\mathbf H}_k^H\right)
\end{eqnarray}
  \item then, with respect to all the remaining free parameters,
  the transmission power for each of
  the $k$-th user is minimized respectively.
 \end{enumerate}
\end{problem}

\subsection{Closed-form optimal solution}\label{subsec:solution}
To solve Problem~\ref{prob:mmse-error}, we employ the inequality
relationship between the trace and determinant of a square matrix
so that the total system error of the MMSE-DFE in
(\ref{mmse-dfe-error}) is lower-bounded by
\begin{eqnarray}\label{equality hold}
{\mathcal
E}\hspace{-0.2cm}&\ge&\hspace{-0.2cm}\frac{1}{N}\sum_{k=1}^K
N_k\det\left({\mathbf
J}^{-1/N_k}_{k}\right)\label{det-matrix-commute}\\
\hspace{-0.2cm}&=&\hspace{-0.2cm}\frac{1}{N}\sum_{k=1}^K N_k
\det\hspace{-0.1cm}\left({\mathbf
I}\hspace{-0.1cm}+\hspace{-0.1cm}{\mathbf T}_k^H{\mathbf
H}_k^H({\boldsymbol\Sigma}_k)^{-1}{\mathbf H}_k{\mathbf T}_k\right)^{\frac{-1}{N_k}}\nonumber\\
\hspace{-0.2cm}&=&\hspace{-0.2cm}\frac{1}{N}\sum_{k=1}^K
N_k\frac{\det({\boldsymbol\Sigma}_k)^{1/N_k}}
{\det({\boldsymbol\Sigma}_{k+1})^{1/N_k}}\nonumber\\
\hspace{-0.2cm}&\ge&\hspace{-0.2cm}\det({\boldsymbol\Sigma}_{K+1})^{-1/N}=2^{-\frac{{\mathcal
I}}{N}}\label{dfe-err-lower-bound}
\end{eqnarray}
Equality in~(\ref{det-matrix-commute}) holds if and only if matrices
${\mathbf J}^{1/2}_{k}$ have equal diagonal R-factors, i.e., in the
DF receiver, the mutual information of the currently detected user
is uniformly distributed over each individual symbol within the
block signal of the user when all the previous user signals have
been perfectly detected. Equality in~(\ref{dfe-err-lower-bound})
holds if and only if $\det({\boldsymbol\Sigma}_k)$ constitutes a
geometrical sequence, i.e.,
\begin{eqnarray}\label{equal ratio}
\left(\frac{\det({\boldsymbol\Sigma}_{1})}
{\det({\boldsymbol\Sigma}_{2})}\right)^{1/N_1}
=\cdots=\left(\frac{\det({\boldsymbol\Sigma}_K)}
{\det({\boldsymbol\Sigma}_{K+1})}\right)^{1/N_K}
\end{eqnarray}
which means the averaged sum mutual information is
uniformly distributed over each 
active subchannel and is equivalent to
\begin{eqnarray}
\det\left({\mathbf J}_{k}\right)=2^{\frac{N_k}{N}{\mathcal I}}
\end{eqnarray}
Therefore, solving Problem~\ref{prob:mmse-error} using the
principle of the Inverse Water-filling solution is finally reduced
to solving the following optimization problem:
\begin{formulation}\label{multiple access throughput}
For any given $K$ non-negative integers $N_1, N_2, \cdots, N_K$
with $N_k\le L_k$, find a sequence of matrices $\{{\mathbf
T}_k\}_{k=1}^K$ such that\hspace{-0.2cm}
\begin{enumerate}
  \item the total power for the $k$th user is minimized subject
  to the constraints that the mutual information for
  User~$k$ is ${\mathcal I}_k=\log\det({\mathbf J}_k)=\frac{N_k}{N}{\mathcal I}$.
  \item within the space of the remaining parameters, (\ref{equality hold})
  holds with equality.
 \end{enumerate}
\end{formulation}
Now, parallel to the result on the inverse water-filling for the
single-user system~\cite{wen08}, we can obtain the following
closed-form solution to Problem~\ref{prob:mmse-error} yielding the
following:
\begin{theorem} Given any $K$ non-negative integers $N_1, N_2, \cdots, N_K$
with $N_k\le L_k$, let
\begin{eqnarray}
{\mathcal A}_k={\mathbf H}_k^H({\boldsymbol\Sigma}_{k})^{-1}{\mathbf
H}_k\qquad {\rm for}\ k=1,2, \cdots, K\nonumber
\end{eqnarray}
and let the eigen value decomposition of ${\mathcal A}_k$ be
${\mathcal A}_k={\mathbf U}_k{\mathbf\Lambda}_k({\mathbf U}_k)^H $
with the diagonal elements in ${\mathbf\Lambda}_k$ arrange in
non-increasing order. Then, the optimal solution to
Problem~\ref{prob:mmse-error} is given by
\begin{eqnarray}
\widetilde{\mathbf T}_k&=&{\mathbf
U}_{N_k,k}({\boldsymbol\Gamma}_k)^{1/2}{\mathbf S}_k,\qquad k=1, 2,
\cdots, K
\end{eqnarray}
where ${\mathbf U}_{N_k,k}$ is the first $N_k$ columns of ${\mathbf
U}_k$, ${\mathbf S}_k$ is an $N_k\times N_k$ unitary matrix denoting
the S-factors of the QRS decomposition of ${\mathbf J}_k^{1/2}$, and
$N_k$ is a pre-assigned subchannel number for the $k$th user. For
the $k$-th user, let $r_k$ be the maximal positive integers such
that
\begin{eqnarray}
\lambda_{n,k}>\left(\frac{\prod_{i=1}^{r_k}\lambda_{i,k}}{2^{{\mathcal
I}_k}}\right)^{1/r_k}\qquad\textrm{for }n=1, 2, \cdots, r_k.
\end{eqnarray}
If $N_k\leq r_k$, the diagonal entries of ${\mathbf\Gamma}_k$ are
determined by
\begin{eqnarray}
\gamma_{n,k}=\left(\frac{2^{{\mathcal
I}_k}}{\prod_{i=1}^{N_k}\lambda_{i,k}}\right)^{1/N_k}-(\lambda_{n,k})^{-1}
\end{eqnarray}
for $n=1, 2, \cdots, N_k$. If $N_k>r_k$, the diagonal entries of
$\Gamma_k$ are assigned by
\begin{eqnarray}
\gamma_{n,k}=\left\{\begin{array}{ll}\left(\frac{2^{{\mathcal
I}_k}}{\prod_{i=1}^{r_k}\lambda_{i,k}}\right)^{1/r_k}-(\lambda_{n,k})^{-1}& n=1,\cdots , r_k\\
0& n=r_k+1,\cdots ,N_k\end{array}\right.\nonumber
\end{eqnarray}
\end{theorem}

\section{SIMULATIONS}
\label{sec:simulation}

In this section, we verify the performance of our optimal
transceiver design using computer simulations. Three examples are
shown in the following:
\begin{enumerate}
\item Fig.~1 shows the scenario of a two-user system. In this
example, each user employs a DMT modulation having 32 subcarriers
and a channel length of $10$. Fig.~1 shows the BER against the sum
mutual information averaged over 100 channel realizations. Three
cases are simulated: $a)$ The number of subcarriers assigned to
User 1 and User 2 is 16 each ($N_k=16, k=1,2$); $b)$ $N_1=16$ and
$N_2=17$, i.e., there is at least one shared subchannel; $c)$
$N_1=17$ and $N_2=17$, i.e., there are at least two subchannels
shared by these two users.

\item Fig.~2 shows a three-user scenario. In this
example, each user employs a DMT modulation having 32 subcarriers.
Again, the channels have a memory size of $10$. Fig.~2 shows the BER
against the sum mutual information averaged over 100 channel
realizations. Three cases are simulated with 100 channel
realizations: $a)$ $N_1=11$, $N_2=11$, and $N_3=10$; $b)$ $N_1=11$,
$N_2=11$, and $N_3=11$, i.e., at least one subchannel is shared;
$c)$ $N_1=12$, $N_2=11$, and $N_3=11$, i.e., at least two
subchannels are shared.

\item Fig.~3 shows a two-user scenario. Here, our
transceiver design is compared with the linear transceiver design
discussed in \cite{multiuser_linear}. In this example, the
simulation environment is the same as Example 1. To ensure a fair
comparison, the sum information and numbers of subcarriers assigned
in our design are calculated from the counterpart algorithm in
\cite{multiuser_linear}. 200 channel realizations are simulated and
taken average over information. Figure~\ref{fig:com_linear} shows
the average bit error rate according to sum mutual information,
while Fig.~4 shows transmission power vs information.
\end{enumerate}

\begin{figure}[htb]\label{fig:2users}
\centering \centerline{\epsfig{figure=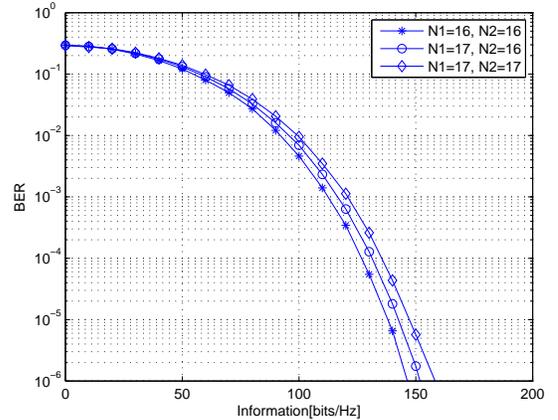,width=8cm}}
  \vspace{-0.5cm}
  \caption{BER vs information in two-user scenario}
\end{figure}

\begin{figure}[htb]\label{fig:3users}
\centering
\centerline{\epsfig{figure=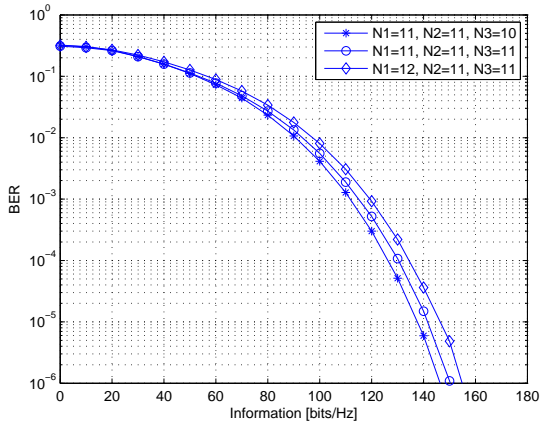,width=8cm}}
  \vspace{-0.5cm}
  \caption{BER vs information in three-user scenario}
\end{figure}

\begin{figure}[htb]\label{fig:com_linear}
\centering \centerline{\epsfig{figure=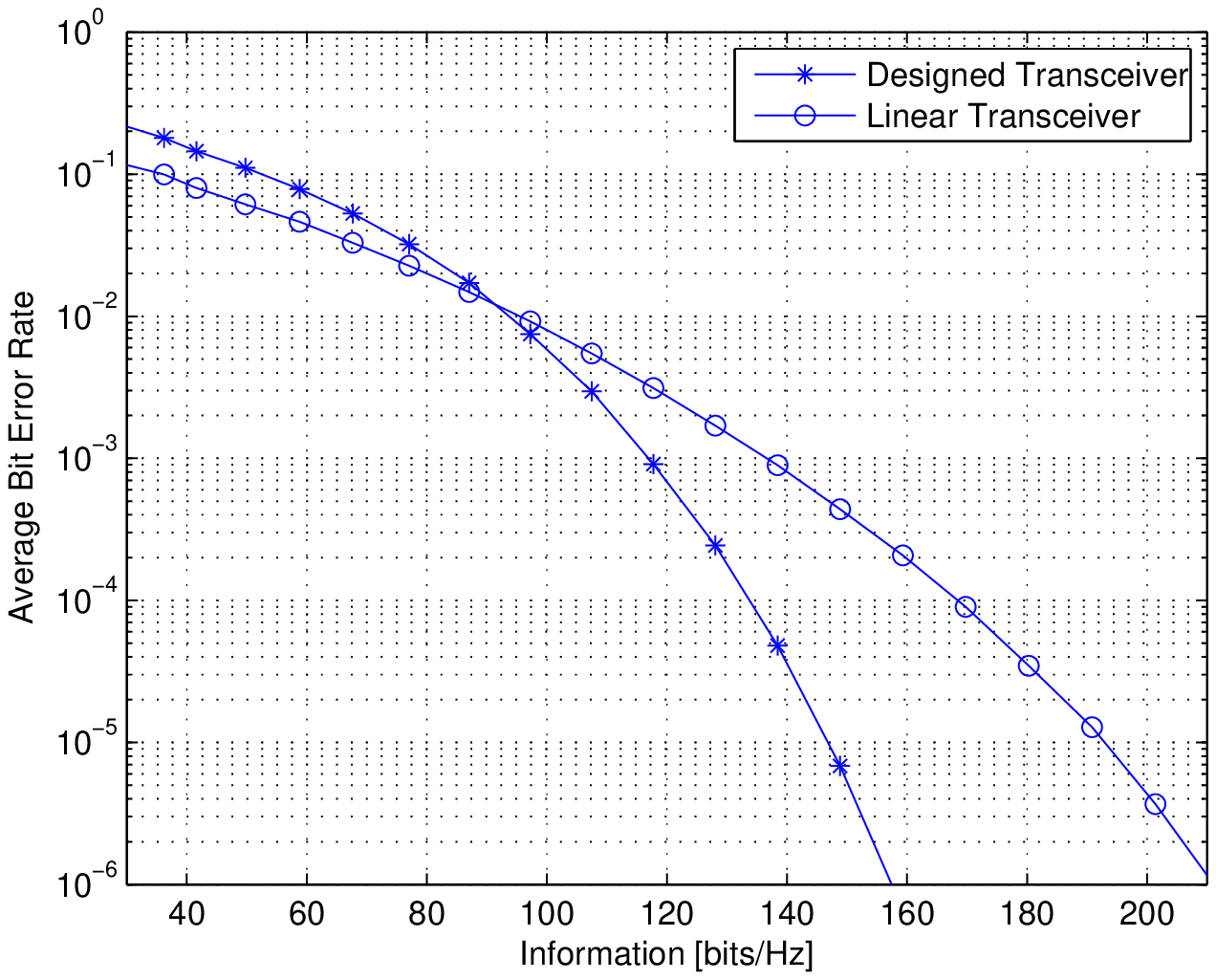,width=8cm}}
  \vspace{-0.3cm}
  \caption{BER vs information: compared with linear MMSE detection}
\end{figure}

\begin{figure}[htb]\label{fig:com_linear_power}
\centering
\centerline{\epsfig{figure=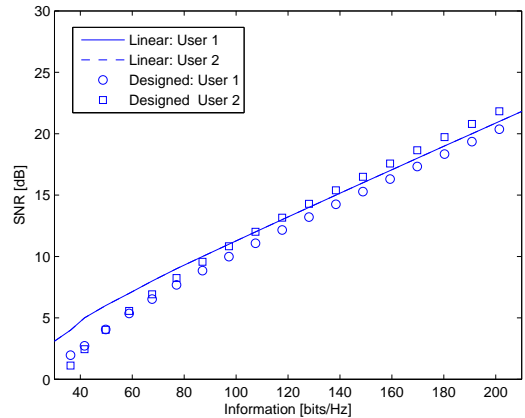,width=8cm}}
  \vspace{-0.5cm}
  \caption{SNR vs information: compared with linear MMSE detection}
\end{figure}

\vspace{-0.5cm}
\section{CONCLUSION}
\label{sec:conclusion}

In this paper, the design of the transceivers for an ISI
multiple-access MIMO communication system using MMSE-DF detection
has been considered. The design goal is to minimize the MSE under
a fixed sum mutual information. The optimal closed-form solution
is obtained when the sum mutual information is uniformly
distributed over each active subchannel and each individual symbol
within the block signal of the user. The latter condition is
achieved by applying the QRS decomposition on the mutual
information matrix of each user, rendering the symbol mutual
information uniformly distributed among all the subcarriers.

\vspace{-0.3cm}

\bibliographystyle{IEEEbib}
\bibliography{wen}

\end{document}